\documentclass[aps,prb,twocolumn,superscriptaddress,amsmath,amssymb]{revtex4-2}

\usepackage{amssymb}
\usepackage{graphicx}
\usepackage{tabularx}
\usepackage{bm}
\usepackage{epsfig}
\usepackage{graphicx}
\usepackage[ansinew]{inputenc}

\begin{document}

\title{Essential aspects of the spontaneous exchange bias effect}

\author{L. Bufai\c{c}al}
\affiliation{Instituto de F\'{\i}sica, Universidade Federal de Goi\'{a}s, 74001-970, Goi\^{a}nia, GO, Brazil}

\author{E. M. Bittar}
\affiliation{Centro Brasileiro de Pesquisas F\'{\i}sicas, 22290-180, Rio de Janeiro, RJ, Brazil}

\date{\today}

\begin{abstract}

Here, we present a review of the phenomenology of the spontaneous exchange bias effect, a phenomenon in which some materials exhibit unidirectional magnetic anisotropy even without the assistance of an external magnetic field applied during its cooling process. We review and discuss the most critical advances in this field of research that flourished more than a decade ago, pointing out its main features as well as its similarities and dissimilarities with the conventional exchange bias effect that has been vastly investigated since the 1950 decade. Finally, we briefly overview the obstacles to advancement in the field and discuss what we believe could be promising roads to overcome them.

\end{abstract}

\maketitle

\section{Introduction}

The exchange bias (EB) effect is a phenomenon of unidirectional magnetic anisotropy (UA) set at the interfaces of different magnetic phases present in heterogeneous systems \cite{Nogues,Rev_Theory,Nogues2}. It is characterized by a shift along the magnetic field ($H$) axis observed in closed curves of magnetization as a function of $H$ [$M(H)$]. Such misplacement of the coercive fields of the $M(H)$ loop is the key to the applicability of this phenomenon in spin valves, magnetic tunnel junctions, and other devices.

The EB effect was discovered more than 60 years ago in a material consisting of ferromagnetic (FM) and antiferromagnetic (AFM) phases \cite{Meiklejohn}. Still, later, it was also shown to occur for AFM-ferrimagnetic (FIM) \cite{AFM-FIM}, FM-FIM \cite{FM-FIM}, FM-spin glass (SG) \cite{FM-SG}, AFM-SG \cite{AFM-SG}, AFM-FIM-SG \cite{Sr2FeCoO6}, etc. For these first discovered materials, the EB phenomenon was ascribed to uncompensated exchange coupling at the interface of the distinct magnetic phases caused by the pinning of some interface spins. For these compounds, as well as for the enormous amount of EB materials developed later, the loop shift could only be achieved after cooling the sample in the presence of an external $H$. Therefore, for decades this cooling field ($H_{FC}$) was understood as an external force necessary to set the UA by breaking the symmetry of the interface moment, \textit{i.e.} by pinning some spins toward the $H_{FC}$ direction (alternatively, the UA could be established in some materials after producing them in the presence of an external field that could lead to an internal remanent magnetization on it).

However, about 13 years ago, the first initially isotropic material exhibiting a robust spontaneous EB effect after being cooled without the presence of an external field was reported \cite{NiMnIn}, opening the path for intense investigation in the research field of the so-called spontaneous EB (SEB) effect (sometimes also called zero-field-cooled EB effect - ZEB). From a scientific viewpoint, there is a great effort to tune the SEB effect and unravel the microscopic mechanisms responsible for it. From the applicability side, the interest in the SEB materials resides in the fact that $H_{FC}$ is no longer necessary to set the UA.

This review compiles the main results of the effort made so far to understand and improve the SEB effect. The text is organized as follows: Section II gives a brief historical background of the conventional EB effect, highlighting some works that can be understood as the seeds of the discovery and development of SEB. Section III focuses on the SEB effect since the early stages of its research, passing through the more important compounds discovered so far, its similarities and disparities with the conventional EB, and the attempts to understand better and tune such effect. Finally, in Section IV, we give a brief outlook of some hindrances to the development of this field of research and the perspectives for the future. This manuscript is mainly focused on (but not limited to) double-perovskite compounds since most SEB materials reported so far belong to this family of compounds.

\section{Historical background: the conventional exchange bias}

\begin{figure}
\begin{center}
\includegraphics[width=0.40 \textwidth]{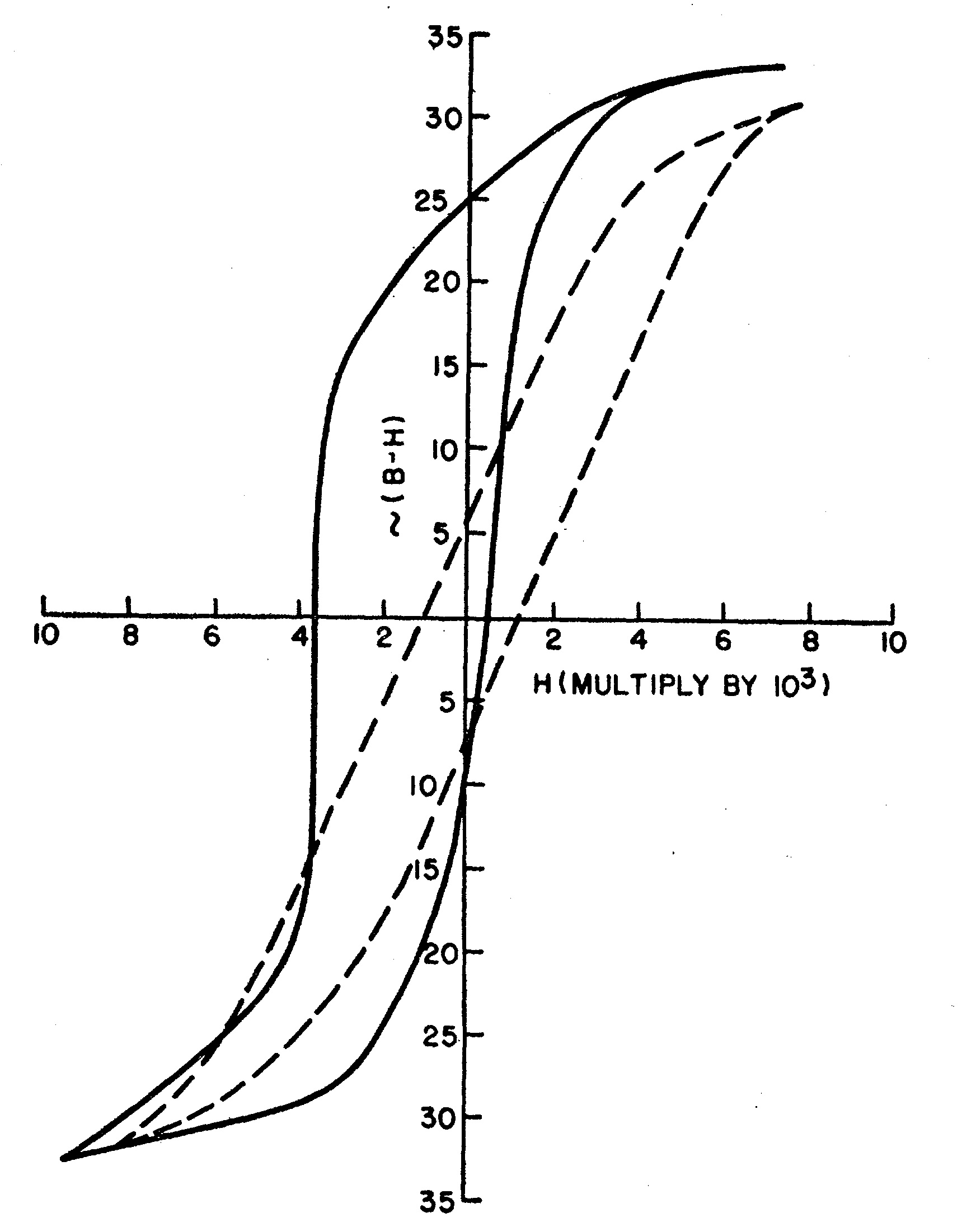}
\end{center}
\caption{M(H) loops of Co@CoO core-shell particles taken at 77 K. The dashed lines show the hysteresis loop when the material is ZFC. The solid lines show the hysteresis loop when the material is cooled with $H_{FC} = 10$ kOe. Extracted from Ref. \onlinecite{Meiklejohn}.}
\label{Fig_Co-CoO_MxH}
\end{figure}

In 1956, W. H. Meiklejohn and C. P. Bean first reported a new type of magnetic anisotropy observed on core-shell particles consisting of an FM core of Co embedded on a CoO AFM shell \cite{Meiklejohn,Meiklejohn2}. The so-called EB effect was manifested by a horizontal shift in the $M(H)$ curve carried after cooling the material in the presence of an external field, resulting from the re-establishment of the ordering of the AFM phase in the presence of the FM phase via a unidirectional interfacial FM-AFM exchange interaction. Conversely, the $M(H)$ loop taken after zero field cool (ZFC) the sample was symmetrical (see Fig. \ref{Fig_Co-CoO_MxH}). 

\begin{figure}
\begin{center}
\includegraphics[width=0.45 \textwidth]{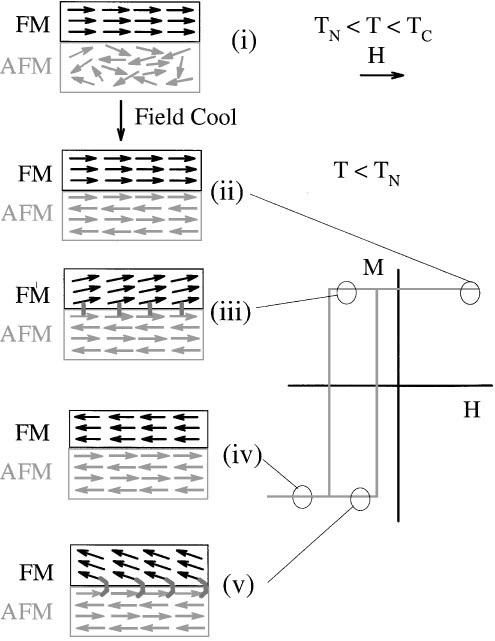}
\end{center}
\caption{Schematic diagram of the spin configuration of an AFM-FM bilayer at different temperatures and stages of the $M(H)$ loop measurement. Note that the spin configurations are just a simple cartoon to illustrate the effect of the coupling, and they are not necessarily accurate portraits of the actual rotation of the FM or AFM magnetizations. Extracted from Ref. \onlinecite{Nogues}.}
\label{Fig_CEB_pinning}
\end{figure}

The EB field, $H_{EB}$ = ($|H_R| - |H_L|$)/2, where $H_R$ and $H_L$ are the coercive fields at respectively the ascending and descending branches of the $M(H)$ curve, gives a measure of the misplacement of the hysteresis loop, which in the case of the fine powders of Co@CoO shown in Fig. \ref{Fig_Co-CoO_MxH} was of approximately -1600 Oe at 77 K \cite{Meiklejohn}. Since its discovery, the exchange anisotropy was qualitatively explained in terms of the pinning of some interface spins toward the $H_{FC}$ direction during the cooling of the material through its N\'{e}el temperature ($T_N$) \cite{Meiklejohn2}. So, most of the theories developed to explain the EB effect invoke an uncompensated AFM interface that pins some nearest-neighbor FM spins via exchange coupling \cite{Nogues,Rev_Theory,Review4}, and naturally the $H_{FC}$  was always considered a necessary condition for the pinning of the interface spins, as schematically depicted in Fig. \ref{Fig_CEB_pinning} for an AFM-FM bilayer. 

Since the EB is primarily an interface effect, it was vastly investigated in thin films consisting of two or more distinct magnetic layers, as well as in core-shell nanoparticles. But this phenomenon is not restricted to these types of materials, being also found in other systems for which different magnetic phases interact, such as inhomogeneous polycrystals, amorphous magnets in contact with ordered magnets, or even distinct types of single crystals put in contact with each other \cite{Nogues,Nogues2}. If, on the one hand, the signature of EB can be easily verified by performing $M(H)$ measurements, on the other hand, the comprehension of the magnetic structure at a material's interface requires much more sophisticated tools. This may be why the attempts to explain the EB effect in most materials were in terms of presumed uncompensated interface spins. However, with the advance of research, it was realized that the EB could occur even for compensated AFM interfaces \cite{Rev_Theory,FeF2-Fe}. Even so, extrinsic factors leading to some small uncompensation in the AFM interface were often considered to explain the EB, such as interface roughness or other defects \cite{interface_roughness1,interface_roughness2}. Consequently, the $H_{FC}$ was still assumed as a necessary condition to set the UA \cite{Koon}, and any shifted $M(H)$ loop observed after ZFC some material was naturally attributed to experimental artifacts such as the presence of remanent field at the equipment during the cooling process or minor loop effects \cite{minor_loop}. Nowadays, the UA set after FC the material is called conventional EB (CEB) effect to distinguish it from the SEB.

Besides the obvious symmetry breaking associated with the UA, some other characteristic features are found in most CEB materials. One of these features is the training effect, consisting of the systematic decrease of $H_{EB}$ as consecutive $M(H)$ loops are measured. Such dependence of the EB effect with the number of hysteresis cycles indicates a metastable configuration at the interfaces. Another mark of CEB is the blocking temperature ($T_B$), above which the EB effect disappears. Naturally, $T_B$ will always be at most $T_N$; in some cases, it is much smaller. For many CEB materials it is observed the enhancement of the coercive field ($H_C$) at temperatures close to $T_B$ \cite{Hc_TB2,Hc_TB1}. Another characteristic feature of the great majority of CEB materials is the negative $H_{EB}$, \textit{i.e.} defining the sign of $H_{FC}$ as positive, the shift of the $M(H)$ curve will be towards negative fields. Here, it is important to mention that there are counterexamples of this feature. Some few materials exhibit positive $H_{EB}$ even after being cooled with $H_{FC}>$ 0 \cite{positiveEB1,positiveEB2}.

\section{The spontaneous exchange bias}

\begin{figure}
\begin{center}
\includegraphics[width=0.45 \textwidth]{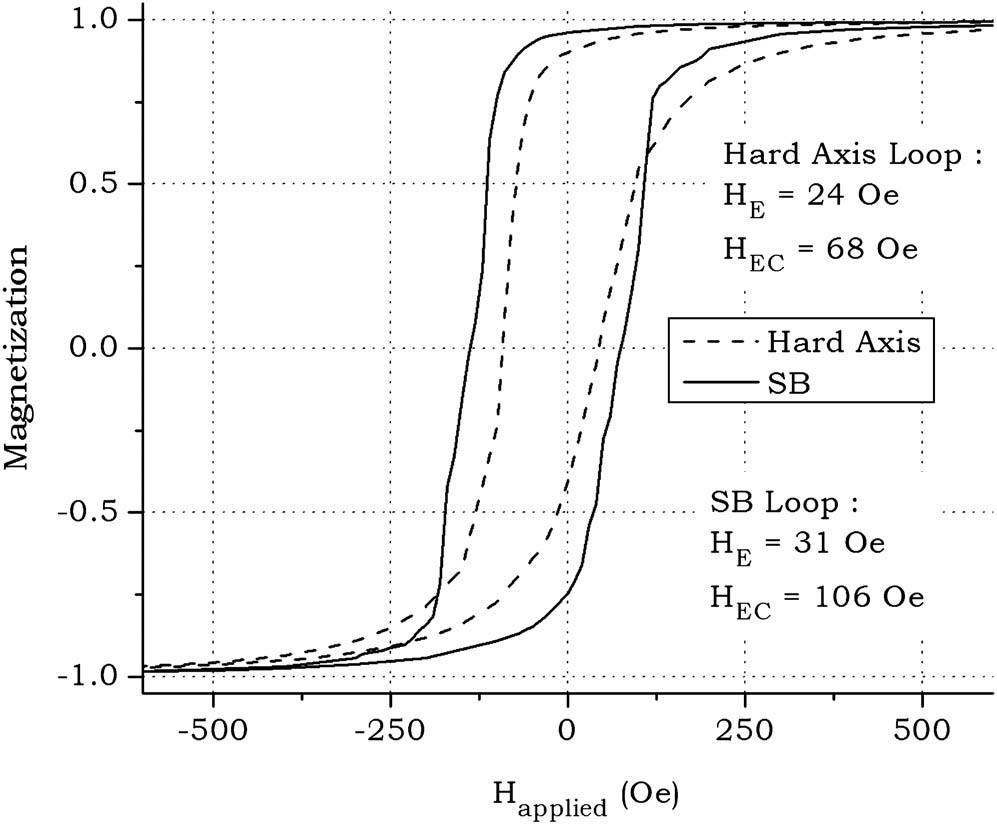}
\end{center}
\caption{Simulations of $M(H)$ curves for the Ni$_{80}$Fe$_{20}$/Ni$_{50}$Mn$_{50}$ system where the Ni$_{50}$Mn$_{50}$ AFM phase presents a biaxial symmetry. The dashed line represents the loop obtained without field cooling, while the solid line shows the loop obtained with $H$ applied along the hard axis.  Extracted from Ref. \onlinecite{Saha}.}
\label{Fig_Saha}
\end{figure}

Although the first observations of the EB effect on fully compensated interfaces, as well as the positive EB, occurred in the context of CEB materials, to some extent, they planted the seeds for the development of SEB by putting in check the necessity of $H_{FC}\neq 0$ to set the UA. At this stage, some few experimental evidence of EB achieved for $H_{FC} = 0$ were reported, albeit not yet discussed in detail \cite{NiFe-CoO,Co1-xMgxO}. But the first clear proposal of a spontaneous exchange anisotropy set after ZFC the system may be attributed to the theoretical study reported in 2007 by J. Saha and R. H. Victora, who proposed a mechanism at which spontaneous UA could be reached in an initially isotropic AFM-FM system, as a consequence of irreversible changes in the energy landscape of the AFM phase caused by the application of $H$ during the $M(H)$ measurement \cite{Saha}. In their work, simulations of $M(H)$ loops for the Ni$_{80}$Fe$_{20}$/Ni$_{50}$Mn$_{50}$ system were performed within the model to show that the first application of $H$ (the so-called virgin curve) was capable to stabilize a preferred direction for some AFM moments, leading to the asymmetric loop depicted in Fig. \ref{Fig_Saha}.

\begin{figure}
\begin{center}
\includegraphics[width=0.45 \textwidth]{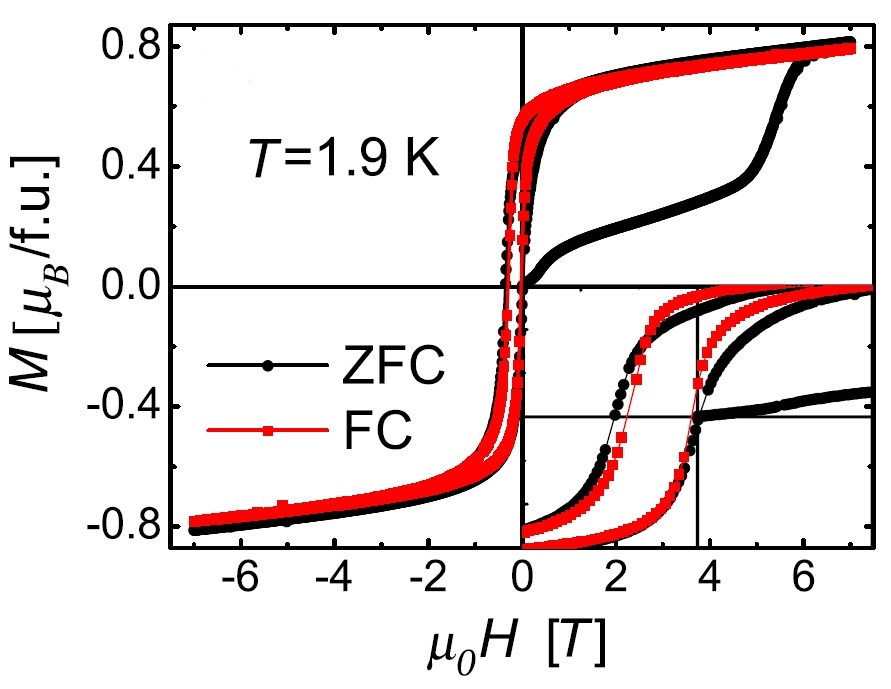}
\end{center}
\caption{$M(H)$ curves measured on Mn$_2$PtGa at 1.9 K. The black curve shows the loop carried after ZFC the sample, while the red one shows the curve taken after cooling with $H_{FC}$ = 70 kOe. The inset shows a magnified view around $H$ = 0, highlighting the asymmetry of the curves.  Adapted from Ref. \onlinecite{Mn2PtGa}.}
\label{Fig_MxH_Mn2PtGa}
\end{figure}

Four years after J. Saha and R. H. Victora's work, B. M. Wang \textit{et al.} reported the first clear realization of robust SEB materials beyond the ambit of experimental artifact, namely the Ni$_{50}$Mn$_{50-x}$In$_{x}$ bulk alloys consisting of an AFM matrix embedding superparamagnetic (SPM) domains that, at lower temperatures, are collectively frozen to form a superspin glass state \cite{NiMnIn}. The ZFC $H_{EB}$ is maxima for $x$ = 13 in this system, being of the order of 1000 Oe at 10 K. Soon later, T. Maity \textit{et al.} reported SEB effect on the BiFeO$_3$-Bi$_2$Fe$_4$O$_9$ nanocomposite \cite{BiFeO3-Bi2Fe4O9}, while A. K. Nayak \textit{et al.} shown a large SEB for the Mn$_2$PtGa Heusler compound (see Fig. \ref{Fig_MxH_Mn2PtGa}) \cite{Mn2PtGa}. At the same time, Z. D. Han \textit{et al.} reported SEB on the off-stoichiometric Ni$_2$Mn$_{1.4}$Ga$_{0.6}$ alloy \cite{NiMnGa}.

\begin{figure}
\begin{center}
\includegraphics[width=0.45 \textwidth]{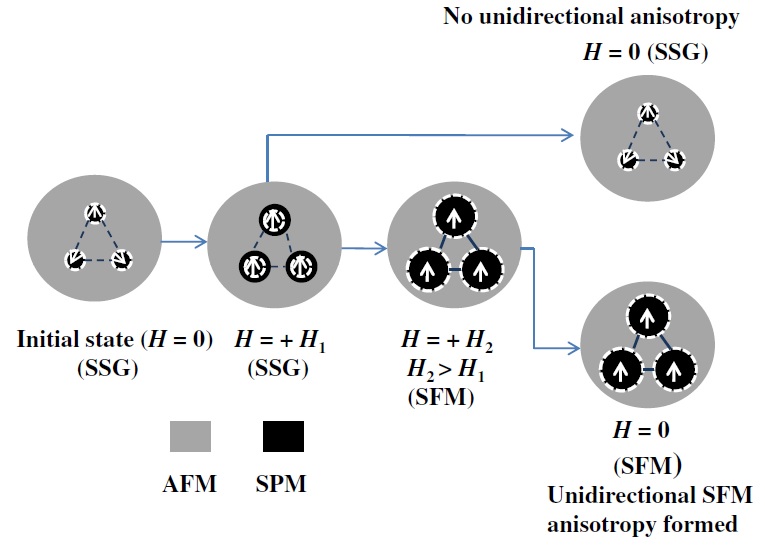}
\end{center}
\caption{Schematic diagram of the evolution of the SPM domains embedded in an AFM single domain on Ni$_{50}$Mn$_{50-x}$In$_{x}$ under the effect of $H$ at temperature below $T_B$. The initial magnetic state after ZFC is an superspin glass state. The white arrows represent the superspin direction of SPM domains. The dashed white circles show the coupling interfaces of SPM and AFM. The dashed (blue) lines represent that the coupling of SPM domains is a glassy coupling, while the solid (blue) lines represent the coupling of SPM domains as an superferromagnetic exchange. Extracted from Ref. \onlinecite{NiMnIn}.}
\label{Fig_diagram_NiMnIn}
\end{figure}

A common ground of the compounds mentioned above is the presence of some glassy magnetism concomitant with other conventional magnetic phases, making these re-entrant spin glass (RSG) systems. Notably, the SG-like phase is a metastable state characterized by a rough energy landscape that can be easily (and irreversibly) altered with temperature and/or magnetic field \cite{Rev_SG}. This fits well with the mechanism proposed by Saha and Victora. Consequently, the SEB of these materials was qualitatively explained in terms of field-induced irreversible changes in the SG-like phase that pins some magnetic moments towards the first applied $H$ direction during the $M(H)$ measurement. Fig. \ref{Fig_diagram_NiMnIn} shows a schematic diagram of the irreversible evolution of the SPM domains on Ni$_{50}$Mn$_{50-x}$In$_{x}$ alloys during the $M(H)$ measurement \cite{NiMnIn}. The application of $H$ increases and aligns the SPM domains (which were initially frozen randomly) to form a superferromagnetic-AFM system. After removing $H$, some superferromagnetic moments will be pinned, leading to the UA observed.

At the same year that the studies on Mn$_2$PtGa, BiFeO$_3$-Bi$_2$Fe$_4$O$_9$ and Ni$_2$Mn$_{1.4}$Ga$_{0.6}$ were reported, the first SEB double-perovskite (DP) appeared, namely the La$_{1.5}$Sr$_{0.5}$CoMnO$_6$ compound \cite{Murthy}. As reported by J. Krishna Murthy and A. Venimadhav, this system is dominated by the Co$^{2+}$--O--Mn$^{4+}$ FM coupling, but the presence of Co$^{2+}$/Co$^{3+}$ and Mn$^{4+}$/Mn$^{3+}$ mixed valences, as well as the anti-site disorder (ASD) at the Co/Mn site (\textit{i.e.} there is a permutation between Co and Mn ions along the lattice), break the long-range FM order to form other interactions such as Co$^{2+}$--O--Co$^{2+}$, Co$^{2+}$--O--Co$^{3+}$, Mn$^{4+}$--O--Mn$^{4+}$, Co$^{2+}$--O--Mn$^{3+}$, etc. As it turns out, the competing exchange interactions lead to the emergence of an RSG state and a giant SEB at lower temperatures. 

In fact, the intrinsic presence of two (or more) transition-metal (TM) ions on double perovskites often leads to disorder and competing magnetic interactions, which are the key ingredients to the emergence of glassy magnetism \cite{Rev_SG}. Thereby, the double perovskite structure was soon recognized as a prospective platform for the growth and investigation of new SEB systems, and it is not a coincidence that most SEB materials discovered so far are double perovskites.

\subsection{The influence of glassy magnetism}

Although the presence of an SG-like phase was clear since the first reported SEB materials, the role played by glassy magnetism on the microscopic mechanism responsible for the UA needed to be clarified at the early research stage. The theoretical work reported by Saha and Victora precedes the experimental verification of SEB. Therefore, the presence of an SG-like phase was not considered on it \cite{Saha}. Even though their model is based on a metastable configuration that evolves with the application of $H$, their scenario is not as drastic as the irreversible changes that usually occur with SG-like systems. Therefore, the spontaneous UA predicted in their work is much smaller than that often found in many SEB compounds. Actually, the fact that the virgin curve lies outside the main $M(H)$ loop for several SEB materials (see Fig. \ref{Fig_MxH_Mn2PtGa}) may be interpreted as a clear indication of the presence of a highly metastable state, where the magnetic ground state is reconfigured with the application of $H$.

Taking into account the fact that the glassy magnetism seems ubiquitous for the great majority of SEB materials, in 2018 L. T. Coutrim \textit{et al.} performed a detailed investigation of its role on the La$_{1.5}$Sr$_{0.5}$CoMnO$_6$ and La$_{1.5}$Ca$_{0.5}$CoMnO$_6$ SEB compounds \cite{modelo_PRB}. The interesting about these double perovskites is that, despite their similar crystal structures and chemical compositions, the former material presents one of the largest $H_{EB}$ reported so far \cite{Murthy}. In contrast, the SEB effect is subtle for the later compound \cite{La1.5Ca0.5}. Since the SG-like state is known for its long-lasting temporal evolution, the authors investigated the time-decay of isothermal remanent magnetization (IRM) curves on these double perovskites to link the temporal evolution of the magnetism with the asymmetry in the $M(H)$ loops. The IRM curves were taken as follows: (i) each material was ZFC down to low temperature ($T$ $<$ $T_B$); (ii) a large $H$ was applied (of the order of the maximum $H$ used in an $M(H)$ curve), and subsequently turned off; (iii) the time-evolution of the remanent magnetization was captured. It is important to note that steps i and ii correspond to the protocol to measure the first quadrant of a typical $M(H)$ loop.

Since the SEB materials are RSG, presenting an SG-like phase as well as other conventional magnetic phases, the IRM curves of La$_{1.5}$Sr$_{0.5}$CoMnO$_6$ and La$_{1.5}$Ca$_{0.5}$CoMnO$_6$ were fitted by adding a term ($M_{conv}$), corresponding to the magnetization of the FM phase, on the stretched exponential equation widely used to describe such curves on canonical SG systems \cite{Rev_SG,stretched1,stretched2}
\begin{equation}
M_{SG}(t)=M_{conv} + M_{0}e^{-\left[(t-t_{1})/t_{p}\right]^{n}}, \label{Eq1}
\end{equation}
where $M_{0}$ represents the pre-exponential magnetization of the SG-like phase, and $t_{p}$ and $n$ ($0<n<1$) are the time and the time-stretch exponential constants, respectively. 

\begin{figure}
\begin{center}
\includegraphics[width=0.45 \textwidth]{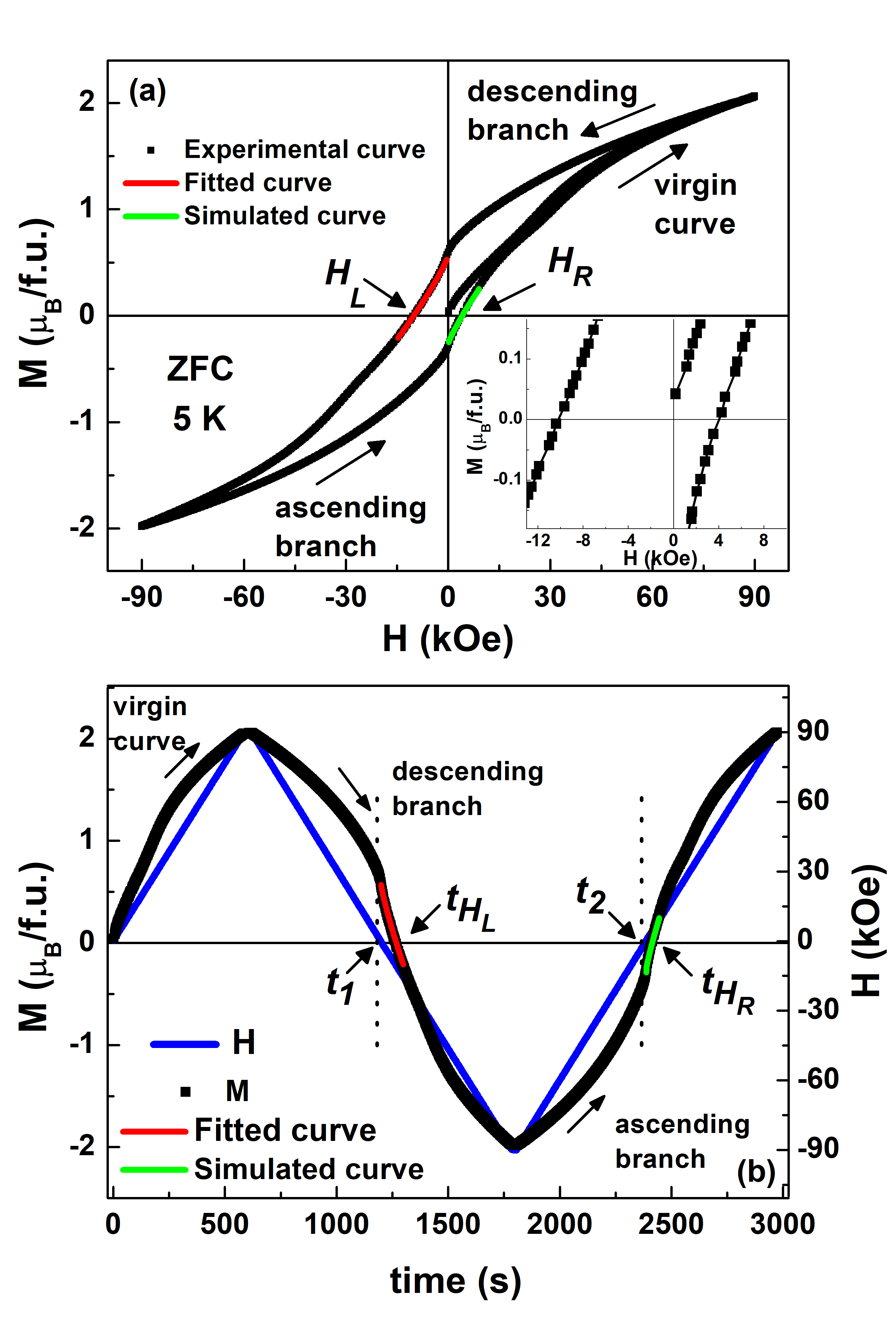}
\end{center}
\caption{(a) $M(H)$ loop of La$_{1.5}$Sr$_{0.5}$CoMnO$_6$ measured after ZFC down to 5 K. The red and green solid lines represent calculated stretches of the curve. The inset shows zoom-in around $M = 0$, evidencing the asymmetry of the curve. (b) The same hysteresis loop is now displayed in $M(t)$ mode. The blue solid line shows $H$ as a function of $t$. Adapted from Ref. \onlinecite{modelo_PRB}.}
\label{Fig_modelo}
\end{figure}

Based on the unusual relaxation of the SG-moments, the authors proposed a phenomenological model for the SEB on double perovskites \cite{modelo_PRB,modelo_JMMM}. From the parameters obtained in the fitting of the IRM curves, together with parameters related to the conventional FM and AFM phases present in the samples, the model could simulate the magnetic behavior of the materials at the regions close to the coercive fields of the $M(H)$ curves. Consequently, the $H_{EB}$ could be calculated. A better understanding of this model can be attained if one reminds that an $M(H)$ loop can be interpreted as a measure of magnetization as a function of time [$M(t)$] since, during the loop measurement, the magnetization ($M$) is captured as a function of $H$, which in turn varies linearly in time ($t$) [(see Figs. \ref{Fig_modelo}(a) and (b)].

In short, the model considers that in the ascending branch of the $M(H)$ loop [\textit{i.e.} the third and fourth quadrants in Fig. \ref{Fig_modelo}(a)], while the material is under the effect of a negative field, some lagger SG-moments are still relaxing from the positive field first applied during the measurement of the virgin curve. These positive moments lead to the decrease of $H_R$, helping to explain the asymmetry in the $M(H)$ curve, which is also contributed by the pinning of some interface spins. However, the fact that the model uses Eq. \ref{Eq1} to account for the evolution of the SG-spins at the second and fourth quadrants of the loop represents an oversimplification of the system. Although at the regions close to $H_R$ and $H_L$ the applied field is relatively small compared to the high fields achieved in a typical $M(H)$ loop, it is not null and may influence the relaxation of the SG-spins significantly. This contrasts with an IRM curve, where the magnetization is captured for $H$ = 0. In any case, the model proposed by L. T Coutrim \textit{et al.} was able to fit the magnetization at the regions of interest and could also explain the remarkably different intensities of the SEB effect observed for the Ca- and Sr-based samples in terms of the distinct relaxation-rates of their SG-phases. Additionally, it could describe the changes in $H_{EB}$ observed for $M(H)$ curves measured with different $H$-sweep rates, as well as the evolution of $H_{EB}$ with temperature and maximum applied field, $H_{max}$ \cite{modelo_PRB,modelo_JMMM}.

Here, it is important to mention that there are some recently reported materials for which a glassy magnetism is not ascribed as a necessary ingredient for the onset of spontaneous UA, many of them presenting near room temperature SEB \cite{Mn3.5Co0.5N,LaMnO3-PbZr0.8Ti0.2O3,BiFeO3-CoFe2O4,BiFeO3-TbMnO3,MnBi,IrMn-FeCo,Mn3Ga,Mn3-xPtxGa,LaFeO3}. For many of these, the UA is claimed to be formed during the initial magnetization process by field-induced pinning of spins at the interface of distinct magnetic phases present in the systems, while for others phase segregation is invoked, with a complex mechanism of spin pinning at the boundaries between distinct phases being used to explain the SEB. We shall return to this subject in subsection D, regarding the SEB dependence on temperature.

\subsection{The training effect}

A common ground of CEB and SEB materials is the training effect, \textit{i.e.} a dependence of their $H_{EB}$ on the number ($n$) of consecutive $M(H)$ measurements. As $n$ increases, $H_{EB}$ systematically decreases to converge to a final value at infinity, $H_{EB}^{\infty}$. This behavior is of fundamental interest for practical applications since it measures the UA's stability, usually related to the relaxation of uncompensated spins after the successive $M(H)$ cycles. For many CEB compounds, the evolution of $H_{EB}$ with $n$ can be well-fitted by the empirical power law equation
\begin{equation}
|{H}_{EB}(n)| = |H^{\infty}_{EB}| + A/\sqrt{n}, \label{EqTE1}
\end{equation}
where $A$ and $H_{EB}^{\infty}$ are adjustable parameters.

\begin{figure}
\begin{center}
\includegraphics[width=0.48 \textwidth]{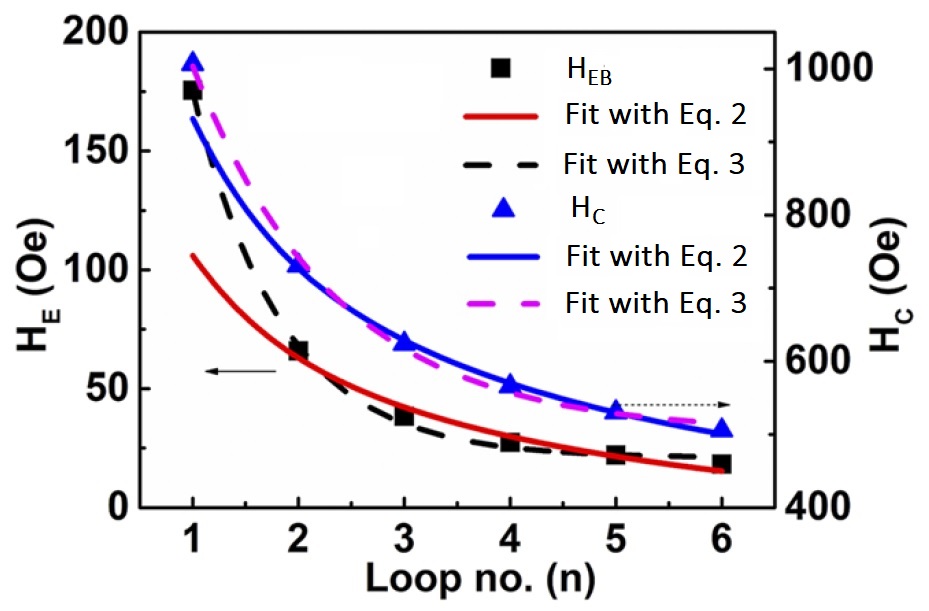}
\end{center}
\caption{Evolution of $H_{EB}$ and $H_C$ with the number of consecutive ZFC $M(H)$ cycles taken at 10 K for a nanocrystalline Mg-ferrite thin film. The solid lines represent the fits with Eq. \ref{EqTE1}, and the dashed ones represent the fits with Eq. \ref{EqTE2}. Adapted from Ref. \onlinecite{Mg-ferrite}.}
\label{Fig_TE}
\end{figure}

The great majority of SEB materials, however, do not fit with Eq. \ref{EqTE1}, especially when the first $M(H)$ loop is considered. Instead, they are better described by a model considering the contributions of both the frozen SG-like spins and the uncompensated rotatable spins at the interfaces \cite{BiFeO3-Bi2Fe4O9,La1.5Ca0.5,JPCM,Mg-ferrite}
\begin{equation}
|{H}^n_{CEB}| = |H^{\infty}_{CEB}| + A_{f}e^{(-n/P_{f})} + A_{r}e^{(-n/P_{r})}, \label{EqTE2}
\end{equation}
where $A_f$ and $P_f$ are parameters associated to the frozen SG-spins, while $A_r$ and $P_r$ are related to the rotatable interface spins. Fig. \ref{Fig_TE} shows the evolution of the SEB effect with $n$ for a nanocrystalline Mg-ferrite thin film, together with the best fittings with Eqs. \ref{EqTE1} and \ref{EqTE2}. Clearly, there is a much better match with the later equation.

Other models were already proposed to explain the training effect on SEB materials \cite{Sm1.5Ca0.5CoMnO6}, but Eq. \ref{EqTE2} is the most widely used. Actually, it was developed before the experimental verification of the SEB effect. It was firstly proposed to describe the evolution of $H_{EB}$ and $H_C$ with $n$ on a CEB film consisting of disordered AFM/FM interfaces, namely the NiFe/IrMn bilayer \cite{NiFe-IrMn}. A shadow can be launched using such an equation since it can be argued that its good matching with the experiments is just a consequence of the fact that it presents several free parameters to be adjusted. 

\subsection{The influence of $H_{FC}$ and $H_{max}$}

For many CEB materials, there is a dependence of $H_{EB}$ with the amplitude of $H_{FC}$, where in general, $|H_{EB}|$ increases with $H_{FC}$ up to a certain value, above which it tends to stabilize independently of $H_{FC}$. However, this is not a universal characteristic of CEB systems. Several compounds do not follow this trend, for instance, the frustrated triangular-lattice AFM Ba$_3$NiIr$_2$O$_9$, whose $H_{EB}$ firstly increases followed by a decrease with increasing $H_{FC}$ \cite{Ba3NiIr2O9}, or the FeF$_2$/Fe bilayer on which $H_{EB}$ changes its sign from negative to positive depending on the intensity of $H_{FC}$ \cite{positiveEB1}.

As for the CEB, the SG state is also very sensitive to its previous thermal and $H$ cycles. It is isotropic after ZFC since the magnetic moments are randomly frozen. Still, it may become anisotropic when cooled at $H_{FC}$ $\neq$ 0, with the applied field favoring the freezing of the spins towards its direction \cite{Rev_SG,H_dependence_SG}. From this, one could expect the enhancement of UA for SEB materials after being cooled with $H_{FC}$ $\neq$ 0. This indeed occurs for several compounds \cite{LaxSm1-xCrO3,CoIr,La1.5Sr0.5Co1-xGaxMnO6,Mn3.5Co0.5N,Pb6Ni9(TeO6)5,La1.5Ba0.5}. The mechanism invoked to explain such behavior is similar to that used to describe the enhancement of $H_{EB}$ with increasing $H_{FC}$ in some CEB materials, for which it is argued that the presence of $H$ already at higher temperatures favors the pinning of some interface spins toward the $H$ direction, thus enhancing the UA.

\begin{figure}
\begin{center}
\includegraphics[width=0.46 \textwidth]{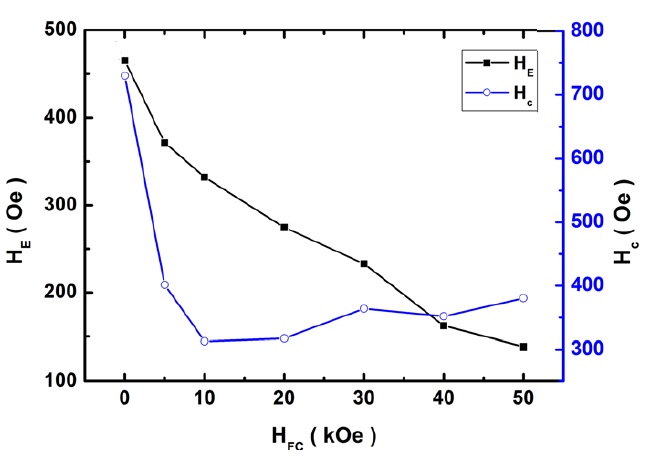}
\end{center}
\caption{Evolution of $H_{EB}$ ($H_E$) and $H_C$ with $H_{FC}$ for the Ni$_2$Mn$_{1.4}$Ga$_{0.6}$ alloy, measured at 10 K. Adapted from Ref. \onlinecite{NiMnGa}.}
\label{Fig_H_FC}
\end{figure}

Nevertheless, there are some SEB materials for which $H_{EB}$ is nearly independent of $H_{FC}$, which is understood as a confirmation that the first application of $H$ during the $M(H)$ measurement (\textit{i.e.} the measurement of the virgin curve) plays for the SEB materials a similar role of the cooling field in the CEB compounds \cite{Mn2PtGa}. There are also some few SEB compounds for which $H_{EB}$ decreases for $H_{FC}$ $\neq$ 0 (see the inset of Fig. \ref{Fig_MxH_Mn2PtGa}) \cite{NiMnIn,NiMnGa}. Fig. \ref{Fig_H_FC} shows the evolution of $H_{EB}$ and $H_C$ with $H_{FC}$ for $M(H)$ loops measured at 10 K on the Ni$_2$Mn$_{1.4}$Ga$_{0.6}$ SEB alloy. The monotonic decrease of $H_{EB}$ with the increase of $H_{FC}$ is interpreted as due to the increase of the FM volume fraction in the spent of the SG phase \cite{NiMnGa}.

\begin{figure}
\begin{center}
\includegraphics[width=0.48 \textwidth]{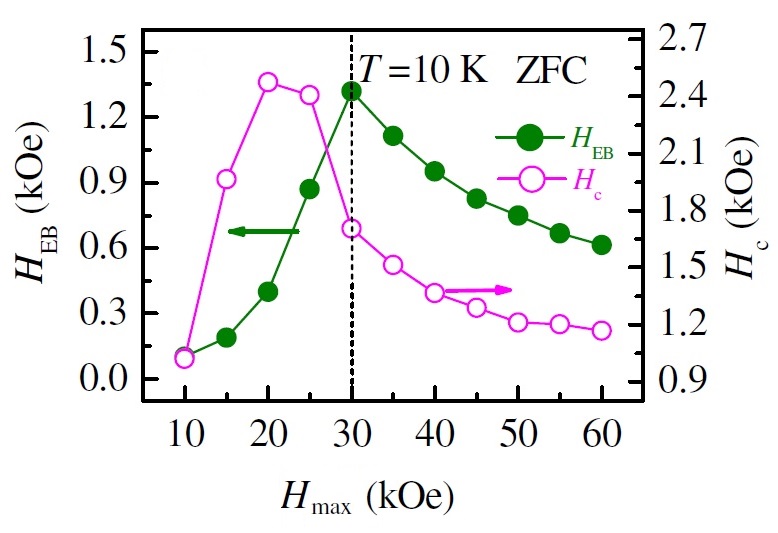}
\end{center}
\caption{$H_{EB}$ and $H_C$ as a function of $H_{max}$ in Ni$_{50}$Mn$_{37}$In$_{13}$ alloy at 10 K, measured after ZFC. The dotted line shows the position of the critical field. Adapted from Ref. \onlinecite{NiMnIn}.}
\label{Fig_Hmax}
\end{figure}

Regarding $H_{max}$, its influence on the SEB effect was scarcely reported up to the present time. But for most of the materials for which such investigation was carried, there seems to be a typical behavior where $H_{EB}$ firstly increases with $H_{max}$ up to a critical value, above which it decreases \cite{NiMnIn,Mn2PtGa,Sm1.5Ca0.5CoMnO6,CoIr2,Ni50Mn38Ga12-xSbx}. Fig. \ref{Fig_Hmax} shows this trend for Ni$_{50}$Mn$_{37}$In$_{13}$ alloy as representative of several compounds for which similar behavior is found. In the case of this alloy, this could be interpreted as due to the fact the increase of $H_{max}$ (below the critical field) induces the growth and stabilization of the FM-like phase present in the system, as well as the exchange interactions at the interfaces, increasing $H_{EB}$ (and $H_C$). But $H_{max}$ values larger than the critical field may change the spin structure in the SG-like phase, \textit{i.e.} it may be strong enough to drag some otherwise pinned spins toward $H$ direction during the field cycle, thus reducing $H_{EB}$.

An exception to this trend is the Sr$_2$FeIrO$_6$-La$_{0.67}$Sr$_{0.33}$MnO$_3$ film, which is one of the few materials claimed to exhibit positive SEB \cite{Sr2FeIrO6-La0.67Sr0.33MnO3,LaMnO3-PbZr0.8Ti0.2O3,SrFeO3-SrCoO3}. In that case, $H_R$ is nearly independent of $H_{max}$ while $|H_L|$ systematically increases with it, leading to the decrease of $H_{EB}$ (and to the increase of $H_C$) with the rise of $H_{max}$.

As aforementioned, for the first reports of the SEB effect, there were doubts about whether the loop shifts were genuinely related to EB or due to experimental artifacts. This is intrinsically linked to $H_{max}$ since, if the applied field is not high enough to overcome the anisotropy field, the asymmetry in the $M(H)$ curve can be associated with minor loop, an effect not related to EB that can be observed on many conventional magnetic materials \cite{minor_loop,Bertotti}. Taking a FM system as an example, the hysteresis observed on its $M(H)$ loop is usually related to domain wall motion. However, for applied fields higher than a critical value (the anisotropy field), the magnetization becomes a single-valued function of $H$, since the pre-existing domain configuration is wiped out. Consequently, the ascending and descending branches of the $M(H)$ curve coincide at the high field region. Conversely, when a field cycle is taken with $H_{max}$ is smaller than this critical value, the resulting hysteresis loop does not encompass the reversible region, usually being unclosed loops, and exhibiting different coercive fields and remanent magnetizations for the ascending and descending field branches of the curve \cite{minor_loop}. In the context of CEB, there were several materials initially claimed to present EB effect, but for which the loop shifts were later on recognized as coming from minor loop effects \cite{SrRuO3,SrRuO3_comment,SrRuO3_erratum}.

\begin{figure}
\begin{center}
\includegraphics[width=0.48 \textwidth]{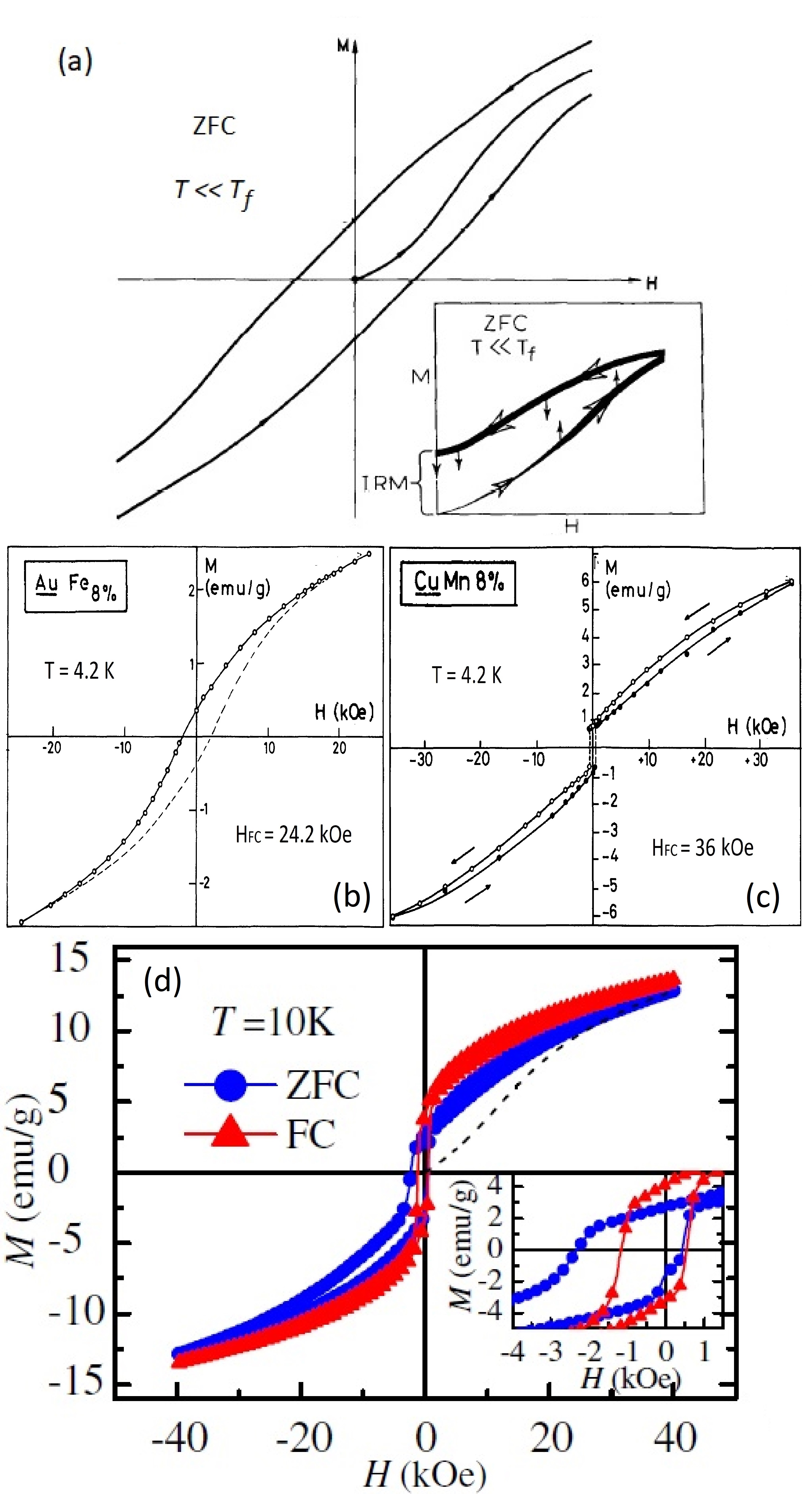}
\end{center}
\caption{(a) Schematic of a typical ZFC hysteresis loop of a SG. The inset shows a magnified view of the high field region, highlighting the lack of saturation. (b) and (c) show $M(H)$ loops of AuFe$_{8\%}$ and CuMn$_{8\%}$ canonical SG alloys, respectively, measured at 4.2 K. (d) ZFC and FC $M(H)$ loops measured at 10 K on the NiMnIn13 SEB material. The inset shows a zoomed-in view of the small field region, highlighting the loop shift. Adapted from Refs. \onlinecite{NiMnIn,Mydosh,CanonicalSG}.}
\label{Fig_minor_loop}
\end{figure}

Here, we must recall the presence of glassy magnetism for the SEB materials. Due to the metastable character of a SG-like state, one may not expect its complete saturation and reversibility on a typical $M(H)$ curve \cite{Mydosh}. Fig. \ref{Fig_minor_loop}(a) shows a schematic ZFC hysteresis loop of a typical SG system, while Figs. \ref{Fig_minor_loop}(b)-(c) show experimental FC $M(H)$ curves for two established canonical SG materials below their freezing temperatures, namely the AuFe$_{8\%}$ and CuMn$_{8\%}$ alloys \cite{CanonicalSG}. For comparison, Fig. \ref{Fig_minor_loop}(d) depicts ZFC and FC $M(H)$ loops measured on the SEB alloy NiMnIn13 at 10 K \cite{NiMnIn}. The characteristics of these curves make it more difficult to unambiguously confirm that the loop shift of an alleged SEB material represents a truly UA. In any case, there are some aspects of the experimental results that can be used to distinguish the loop shift of a true SEB from that associated with a minor loop: the $H_{max}$ must be high enough to ensure that the loop is closed, and the ascending and descending branches of the curve must reasonably coincide at high fields. Although these features alone are not enough to undoubtedly determine that a loop shift is due to EB, they strongly indicate it.

\subsection{The influence of temperature}

\begin{figure}
\begin{center}
\includegraphics[width=0.48 \textwidth]{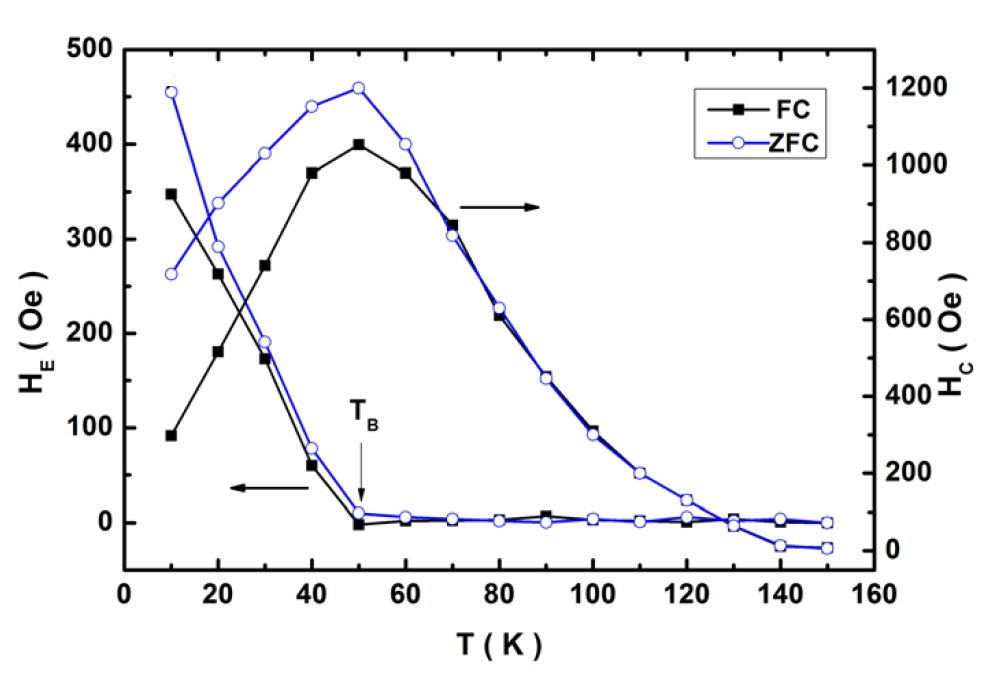}
\end{center}
\caption{Evolution of $H_{EB}$ ($H_E$) and $H_C$ with temperature for the Ni$_2$Mn$_{1.4}$Ga$_{0.6}$ alloy, with M(H) loops measured after ZFC (open circles) and after FC with $H_{FC}$ = 10 kOe. The lines are guides for the eye. Adapted from Ref. \onlinecite{NiMnGa}.}
\label{Fig_TB}
\end{figure}

Making an analogy with the effect of increasing temperature on several CEB materials, one could expect for the SEB systems that the thermal energy may allow some pinned spins to overcome the energy barrier of the exchange interactions in a way to flip toward $H$ direction during the $M(H)$ measurement, leading to the decrease of $H_{EB}$ with increasing temperature, until it vanishes at the blocking temperature, $T_B$. Indeed, this occurs for many SEB materials \cite{Mn2PtGa,Murthy,La1.5Ca0.5,JPCM,Mg-ferrite,Sm1.5Ca0.5CoMnO6,Pb6Ni9(TeO6)5,Ni50Mn38Ga12-xSbx,La0.7Sr0.3MnO3-NiO}, as exemplified in Fig. \ref{Fig_TB} for the Ni$_2$Mn$_{1.4}$Ga$_{0.6}$ alloy \cite{NiMnGa}. However, there are compounds for which the thermal energy leads to more complex reconfiguration of the magnetic phases, resulting in non-monotonic changes of $H_{EB}$ with temperature \cite{BiFeO3-Bi2Fe4O9,LaxSm1-xCrO3,Mn3.5Co0.5N,La1.5Ba0.5,BiFeO3,SmFeO3}. 

\begin{figure}
\begin{center}
\includegraphics[width=0.48 \textwidth]{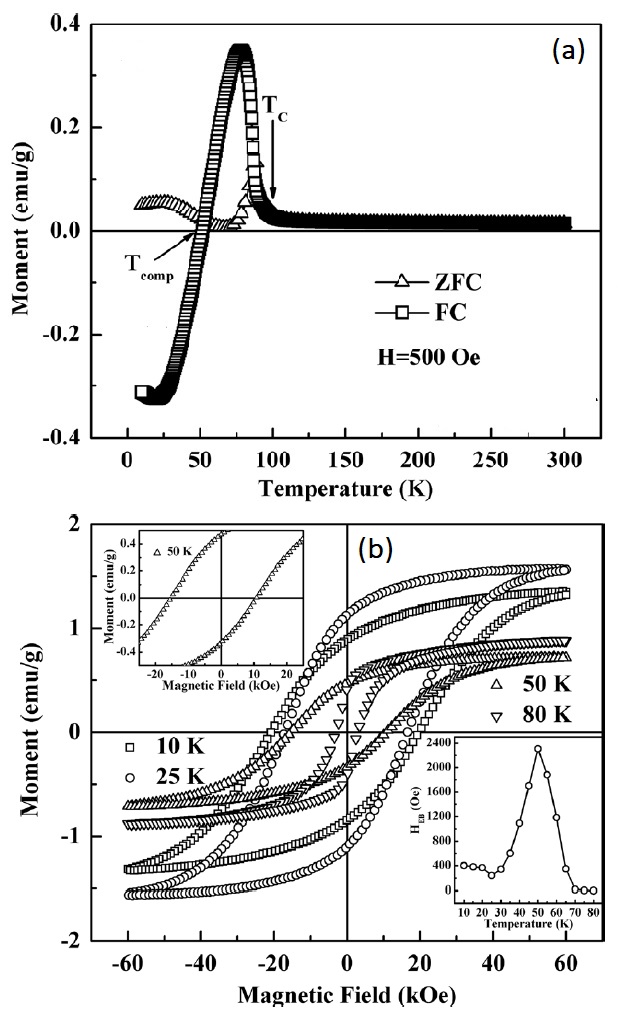}
\end{center}
\caption{(a) ZFC-FC Magnetization as a function of temperature for the Co$_{0.8}$Cu$_{0.2}$Cr$_2$O$_4$ SEB oxide, measured with $H$ = 500 Oe. (b) ZFC $M(H)$ curves were measured at different temperatures. The upper inset shows a magnified view of the 50 K curve, highlighting the loop shift. The bottom inset shows the evolution of $H_{EB}$ with temperature. The lines are guides for the eye. Adapted from Ref. \onlinecite{Co0.8Cu0.2Cr2O4}.}
\label{Fig_Tcomp}
\end{figure}

In particular, some SEB materials presenting compensation temperature, \textit{i.e.} their magnetization as a function of temperature curves show inversion of the magnetization sign at a particular temperature ($T_{comp}$), exhibit a dependence of $H_{EB}$ with $T_{comp}$ \cite{La1.5Sr0.5Fe0.4Co0.6MnO6,Co0.8Cu0.2Cr2O4}. Fig. \ref{Fig_Tcomp}(b) illustrates this situation for the Co$_{0.8}$Cu$_{0.2}$Cr$_2$O$_4$ SEB oxide, showing the increase of $H_{EB}$ with decreasing temperature down to 50 K, when the magnetization switches its sign and $H_{EB}$ starts to decrease. This behavior was interpreted as follows: below $T_{comp}$, the ''negative" magnetic moments that start to develop partially compensate the ''positively" pinned spins, leading to the decrease of $H_{EB}$ \cite{Co0.8Cu0.2Cr2O4}.

Since an SG-like phase is a common ground of many SEB materials, it is natural to expect the spontaneous UA to be set at temperatures below the emergence of glass magnetism on these. This is indeed the case for the great majority of SEB compounds. Nevertheless, there are some materials reported to exhibit SEB effect persisting up to very high temperatures \cite{BiFeO3-Bi2Fe4O9,Mn3.5Co0.5N,LaMnO3-PbZr0.8Ti0.2O3,BiFeO3,BiFeO3-CoFe2O4,BiFeO3-TbMnO3,MnBi,IrMn-FeCo,Mn3Ga,Mn3.1Sn0.9}. Here is essential to mention that for some of these compounds the freezing temperature of the SG-like phase was not investigated in detail (or the presence of glassy magnetism was not even verified). For others, the ordering temperatures of the conventional magnetic phases were also not adequately determined, launching doubts about whether the $M(H)$ measurements came after a truly ZFC protocol. In other cases, the shift in the $M(H)$ curve is very small, opening the possibility of it being related to instrumental artifacts such as a trapped field in the magnet used for the experiments. There are also reports for which it is unclear if the $H_{max}$ used in the experiments was high enough to reach unambiguously closed $M(H)$ loops in a way that the asymmetry in the curves could be related to minor loop effects.

Regarding the $H_C$ dependence with temperature in SEB systems, a local maximum is usually observed roughly at the same temperature interval where $H_{EB}$ decreases, as depicted in Fig. \ref{Fig_TB}. This is analogous to the behavior of several CEB systems, as commented in Section II, and is generally ascribed to thermal energy-induced disentanglement of some previously pinned spins that are now able to rotate with $H$, thus increasing $H_C$. But again, this is not a universal feature of SEB; some materials do not follow this trend \cite{BiFeO3-Bi2Fe4O9,La1.5Ca0.5,Sr2FeIrO6-La0.67Sr0.33MnO3,La0.7Sr0.3MnO3-NiO,SmFeO3}.

\subsection{Cationic disorder}

As aforementioned, the EB effect is intrinsically related to the presence of distinct magnetic phases in heterogeneous systems \cite{Nogues}. Analogously, the glassy magnetism ever-present on SEB materials is characterized by disorder and competing magnetic phases \cite{Rev_SG}. Therefore, a natural step in the research of the SEB effect is investigating the influence of disorder on the spontaneous UA formed. In this regard, the double perovskite oxides with general formula $A_2$BB'O$_6$, where $A$ is usually a rare-earth or alkaline-earth ion and B/B' are distinct TM ions, represent a prospective study case because they almost always exhibit a significant ASD, which might lead to competing magnetic interactions and frustration \cite{Serrate,Sami,Rev_ASD}. 

A feasible way to tune the ASD on double perovskites is by performing chemical substitution at the $A$ or B/B' sites with ions presenting different radii and/or valence states. For the La$_{2-x}A_x$CoMnO$_6$ ($A$ = Ca, Sr) series, for instance, it was found a direct relation between the evolution of SEB along both series and the doping-induced changes in ASD at the Co/Mn site \cite{Murthy2,La2-xCax}. This was understood by the fact that the Ca$^{2+}$/Sr$^{2+}$ doping at the La$^{3+}$ site in La$_2$CoMnO$_6$ affects the ASD by the ionic radii mismatch, and also induces mixed valence states at the TM ions sites, Co$^{2+}$/Co$^{3+}$ and Mn$^{4+}$/Mn$^{3+}$. Both the ASD and mixed oxidation states are responsible for interrupting the Co$^{2+}$-O-Mn$^{4+}$ long-range FM order by introducing secondary magnetic interactions. This gives rise to the RSG behavior observed on both series at low temperatures.

Introducing a third magnetic element can also tune the SEB by enhancing the magnetic disorder. For example, a partial substitution of Co by Fe on the La$_{1.5}$Sr$_{0.5}$CoMnO$_6$ leads to additional exchange interactions between the TM ions, some of which are uncompensated, resulting in a SEB effect that can invert its sign depending on the doping concentration \cite{La1.5Sr0.5Fe0.4Co0.6MnO6,La1.5Sr0.5Co1-xFexMnO6}. Interestingly, introducing non-magnetic ions at the B/B' sites can also enhance the SEB. In the case of the La$_{1.5}$Sr$_{0.5}$Co$_{1-x}$Ga$_x$MnO$_6$ series, the incorporation of a small concentration of Ga$^{3+}$ (3$d^{10}$) along the lattice leads to changes in the ASD and in the SG-like state that enhance the SEB \cite{La1.5Sr0.5Co1-xGaxMnO6}.

Alternatively, the SEB effect can also be achieved by introducing magnetic rare-earth elements at the $A$-site. As an example, LaCrO$_3$ shows no spontaneous UA, but the partial substitution of La$^{3+}$ by Sm$^{3+}$ results in additional internal fields caused by the exchange interactions between Sm and Cr that set the SEB effect, whose magnitude increases with the Sm content \cite{LaxSm1-xCrO3}.  Similar pictures were drawn to explain the SEB in other perovskites on which magnetic rare-earth elements are present \cite{Sm1.5Ca0.5CoMnO6,SmFeO3,Er2CoMnO6,Pr2-xSrxCoMnO6,Y0.95Eu0.05MnO3}.

The investigation of the effect of doping in the context of SEB is not confined to perovskites. A. K. Nayak \textit{et al.} found that the uncompensated AFM coupling between Mn ions in the Mn$_{3-x}$Pt$_x$Ga Heusler alloys are influenced by the Pt-concentration, thus affecting the SEB \cite{Mn2PtGa,Mn3-xPtxGa}. In the Ni$_{50}$Mn$_{38}$Ga$_{12-x}$Sb$_x$ alloys, the SEB changes with Sb-content as the glassy magnetic domains evolve, and $H_{EB}$ is maxima for $x$ = 2 when the system is in a phase boundary between canonical and cluster SG \cite{Ni50Mn38Ga12-xSbx}.

\subsection{The influence of crystal structure}

The crystal structure plays its role in the SEB effect not only through ASD since it also strongly affects the exchange interactions between the magnetic ions. In the case of perovskites, the structural distortions directly impact the tilting and rotation of the oxygen octahedra surrounding the TM ions, thus affecting its environmental crystal electric field and the hybridization between the magnetic ions, which is usually a superexchange interaction mediated by the intervening oxygen ions. In a study of the La$_{1.5}A_x$CoMnO$_6$ ($A$ = Ba, Ca, Sr) double perovskites, L. T. Coutrim \textit{et al.} proposed that the structural changes caused by altering the $A$ element play an important role on the material's magnetic properties by changing the Co$^{3+}$ spin state, which affects the system's SEB through the uncompensated Co-O-Mn coupling \cite{PRB2019}. A similar mechanism was adopted to explain the evolution of $H_{EB}$ with Ba-content on the La$_{1.5}$(Sr$_{0.5-x}$Ba$_x$)CoMnO$_6$ series \cite{APL}. The lattice expansion caused by the Ba to Sr substitution increases the fraction of high spin Co$^{3+}$, which, on the one hand, strengthens the AFM couplings, acting to increase the SEB. On the other hand, the increased portion of high spin Co$^{3+}$ acts to decrease the uncompensation in the Co-O-Mn coupling. These competing effects lead to the initial increase of $H_{EB}$ with doping, followed by a decrease, with a very large SEB being observed for the intermediate Sr/Ba region.

In the context of the La$_{2-x}$Ca$_x$CoMnO$_6$ series, differently than the scenario drawn in Ref. \onlinecite{La2-xCax}, at which the evolution of the SEB effect was directly associated with changes in the ASD, J. R. Jesus \textit{et al.} interpreted the increase of $H_{EB}$ with Ca-doping by the strengthening of the uncompensated magnetic coupling at the magnetic interfaces caused by the enhanced AFM phase, which in turn is related to the system's crystal structure \cite{meu_La2-xCax}.

In thin films, the SEB effect is strongly impacted by strain. As an example, for the La$_{0.67}$Sr$_{0.33}$MnO$_3$/PbZr$_{0.8}$Ti$_{0.2}$O$_3$/La$_{0.67}$Sr$_{0.33}$MnO$_3$ sandwich, AFM islands are formed in the La$_{0.67}$Sr$_{0.33}$MnO$_3$ layers due to large strain imposed through the PbZr$_{0.8}$Ti$_{0.2}$O$_3$ layer. Changing the PbZr$_{0.8}$Ti$_{0.2}$O$_3$ thickness alters the strain, affecting the SEB \cite{LaMnO3-PbZr0.8Ti0.2O3}. Similarly, in the Si/Pt/Ni$_{45}$Mn$_{55}$/Co/Pt thin films, the exchange coupling between Ni$_{45}$Mn$_{55}$ and Co layers is particularly sensitive to the thickness of the Pt buffer layer, the SEB being only observed for thicker Pt layers \cite{Co-NiMn-filme}.

Interestingly, an IrMn/FeCo bilayer fabricated by magnetron sputtering shows an SEB effect driven by a structural phase transition in the IrMn AFM layer \cite{IrMn-FeCo}. Immediately after deposition, the sample shows no EB. But when the system is let to relax in time, under the effect of the remanent magnetization present on the FeCo FM layer, a structural transition starts to develop on the IrMn AFM layer, after which the material exhibits the SEB effect.

\subsection{The influence of grain size}

The role of grain size on the UA was vastly investigated in CEB polycrystals such as nanoparticles and core-shell systems \cite{Nogues2}. Being the CEB an interface effect, at first glance, one could expect a direct relation between the grain's surface area and the UA. However, the situation is not that simple because the grain size substantially affects other essential parameters in the context of EB. Consequently, $H_{EB}$ increases with the grain size for some materials while the opposite trend is observed for other systems \cite{Nogues,Nogues2}.

\begin{figure}
\begin{center}
\includegraphics[width=0.46 \textwidth]{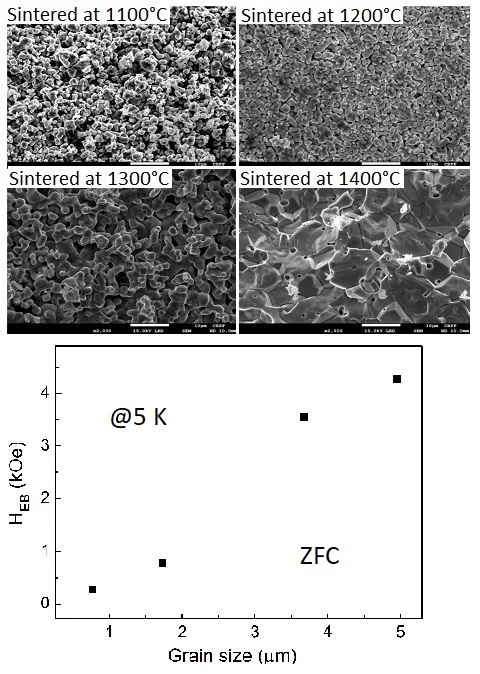}
\end{center}
\caption{(a) Scanning electron microscopy images of the grains of La$_{1.5}$Sr$_{0.5}$CoMnO$_6$ polycrystals produced at different sintering temperatures. (b) Evolution of the ZFC $H_{EB}$ as a function of the average grain size, for $M(H)$ curves measured at 5 K. Adapted from Ref. \onlinecite{Calazans}.}
\label{Fig_grain_size}
\end{figure}

Contrasting with the large number of studies in CEB compounds, the effect of grain size was seldom investigated in SEB materials. Apart from one or other brief qualitative comments on manuscripts that were mainly focused on other SEB features, the only detailed investigation on this subject was performed by C. Macchiutti \textit{et al.} \cite{Calazans}. In this work, polycrystalline samples of the La$_{1.5}$Sr$_{0.5}$CoMnO$_6$ SEB double perovskite sintered at different temperatures resulted in materials with distinct average grain sizes and morphology, leading to different ZFC $H_{EB}$, as depicted in Fig. \ref{Fig_grain_size}. The systematic decrease in the fraction of the SG phase with decreasing the surface-to-core ratio, together with the author's previous work showing that a single crystalline sample of La$_{1.5}$Sr$_{0.5}$CoMnO$_6$ does not show SG-like behavior (nor SEB effect) \cite{single_crystal}, lead them to conclude that the glassy magnetic phase of these polycrystals must reside in the grain boundaries. The increase of $H_{EB}$ with grain size was ascribed to the strengthened exchange interactions between adjacent grains, caused by the facetated morphology of the larger grains.

To some extent, one can make an analogy between the grain size of polycrystalline bulk materials and the layer thickness on thin films. In Sr$_2$FeIrO$_6$/La$_{0.67}$Sr$_{0.33}$MnO$_3$ layers deposited on SrTiO$_3$ substrate, SEB is only observed for layers with a minimum thickness ($\sim$ 10 nm) \cite{Sr2FeIrO6-La0.67Sr0.33MnO3}. For SrFeO$_{3-x}$/SrCoO$_{3-x}$ SEB films grown with a thick non-magnetic SrTiO$_3$ spacer between the SrFeO$_{3-x}$ and SrCoO$_{3-x}$ layers, the FM exchange coupling at the interface continuously weakens with increasing the SrTiO$_3$ thickness, thus reducing $H_{EB}$ \cite{SrFeO3-SrCoO3}. Similar behavior is found for La$_{0.67}$Sr$_{0.33}$MnO$_3$/PbZr$_{0.8}$Ti$_{0.2}$O$_3$/La$_{0.67}$Sr$_{0.33}$MnO$_3$ films, but here the SEB is attributed to strain effects caused by the lattice mismatch between La$_{0.67}$Sr$_{0.33}$MnO$_3$ (LSMO) and PbZr$_{0.8}$Ti$_{0.2}$O$_3$ (PZT), which in turn induces AFM behavior on LSMO. Above some PZT thickness, the contact between the LSMO layers disappears, and so does the SEB \cite{LaMnO3-PbZr0.8Ti0.2O3}. Conversely, in the case of Si/Pt/Ni$_{45}$Mn$_{55}$/Co/Pt thin films, the SEB is only observed for thicker Pt buffer layers. Here, the UA is believed to come from the magnetic texture on the NiMn layer, which is achieved when the Pt buffer layer is introduced \cite{Co-NiMn-filme}.

\subsection{Other ZFC EB systems}

\begin{figure}
\begin{center}
\includegraphics[width=0.49 \textwidth]{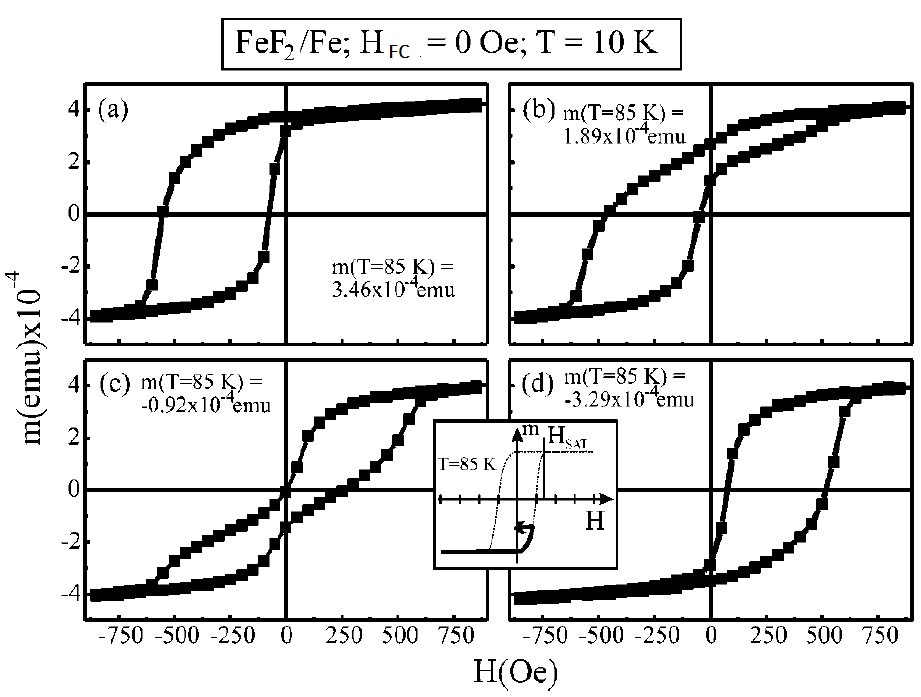}
\end{center}
\caption{$M(H)$ loops measured at 10 K on FeF$_2$/Fe bilayer after ZFC from 85 K in different magnetization states: (a) $m$ = +3.46 $\times$ 10$^{-4}$ emu, (b) $m$ = +1.89 $\times$ 10$^{-4}$ emu, (c) $m$ = -0.92 $\times$ 10$^{-4}$ emu, (d) $m$ = -3.29 $\times$ 10$^{-4}$ emu. The inset shows the procedure to set up different magnetization states at 85 K. The lines are guides to the eye. Adapted from Ref. \onlinecite{tuning_EB}.}
\label{Fig_FeF2}
\end{figure}

Besides the SEB materials described in the previous sections, there are some compounds produced in the presence of $H$ that exhibit the ZFC EB effect. This is the case, for instance, of the polycrystalline Cu/Ni$_{80}$Fe$_{20}$/Mn$_{83}$Ir$_{17}$/Co/AlO$_x$ AFM-FM layer system growth in the presence of in-plane $H$, for which a clear ZFC EB is observed, whereas there is no UA when the film is prepared without $H$ \cite{Ni80Fe20-Mn83Ir17}. The UA can also be set after ZFC when the material's FM component presents remanent magnetization from above $T_N$ \cite{CoO-Ni81Fe19,tuning_EB}. This is shown in Fig. \ref{Fig_FeF2} for AFM-FM FeF$_2$/Fe bilayer samples prepared using electron beam deposition \cite{tuning_EB}. The samples were cooled down to 10 K with $H_{FC}$ = 0, but before the ZFC procedure, they were magnetized with different $H$ at 85 K, \textit{i.e.} at $T$ $>$ $T_N$. As can be seen, the different magnetization states achieved by the different applied $H$ lead to distinct $H_{EB}$.

However, these materials should not be confused with the SEB ones discussed in this review because, in essence, the mechanisms responsible for their ZFC EB can be understood as situations where the materials are passively field-cooled. They present some remanent magnetization before the $M(H)$ measurement, being thus fundamentally different from the initially isotropic SEB materials described in previous sections.
 
\section{Conclusions and outlook}

In conclusion, we have reviewed the SEB properties of several types of materials. Many of these features are similar to those observed in CEB systems, while others are particular to the SEB compounds. Despite the substantial progress made in recent years in elucidating some qualitative details concerning this phenomenon, a complete understanding of this effect requires additional studies, encompassing more sophisticated control of the sample preparation and of the experiments, as well as more detailed microscopic models to explain the physical properties of these systems.

At this stage, it is observed that the great majority of SEB materials present some kind of glassy magnetism. From the point of view of technological application, this represents a clear hindrance since the SG-like behavior usually manifests at low temperatures. Here, the design of materials presenting glassy magnetism above room temperature is desirable. Another possible route to overcome this obstacle could be the development of heterogeneous materials presenting metastable magnetic phases other than the SG-like state, such as the Griffiths phase, which is commonly observed at higher temperatures \cite{Griffiths,Griffiths2}.

For some of the materials claimed to exhibit SEB at room temperature, a more detailed verification of whether the asymmetry on their $M(H)$ curves represents a true EB effect or if it is related to experimental artifacts is necessary. For those whose spontaneous UA is (or will be) confirmed, the next step could be investigating possible ways to tune the EB effect in order to achieve values demanded by specific devices. Again, this requires a more profound understanding of these systems' microscopic mechanisms responsible for the SEB.

\begin{acknowledgements}
This work was supported by the Brazilian funding agencies: Funda\c{c}\~{a}o Carlos Chagas Filho de Amparo \`{a} Pesquisa do Estado do Rio de Janeiro (FAPERJ) [Nos. E-26/202.798/2019 and E-26/211.291/2021], Funda\c{c}\~{a}o de Amparo \`{a}  Pesquisa do Estado de Goi\'{a}s (FAPEG) and Conselho Nacional de Desenvlovimento Cient\'{\i}fico e Tecnol\'{o}gico (CNPq) [Nos. 400633/2016-7, 310889/2020-0 and 305394/2023-1].
\end{acknowledgements}


\begin{thebibliography}{99}

\bibitem{Nogues} J. Nogu\'{e}s and I. K. Schuller, J. Magn. Magn. Mater. 192, 203 (1999).

\bibitem{Rev_Theory} M. Kiwi, J. Magn. Magn. Mater. 234, 584-595 (2001).

\bibitem{Nogues2} J. Nogu\'{e}s, J. Sort, V. Langlais, V. Skumryev, S. Suri\~{n}ach, J. S. Mu\~{n}oz, M. D. Bar\'{o}, Phys. Rep. 422, 65-117 (2005).

\bibitem{Meiklejohn} W. H. Meiklejohn and C. P. Bean, Phys. Rev. \textbf{102}, 1413 (1956).

\bibitem{AFM-FIM} P. J. van der Zaag, R. M. Wolf, A. R. Ball, C. Bordel, L. F. Feiner, R. Jungblut, J. Magn. Magn. Mater. \textbf{148}, 346-348 (1995).

\bibitem{FM-FIM} William C. Cain and Mark H. Kryder, J. Appl. Phys. \textbf{67}, 5722-5724 (1990).

\bibitem{FM-SG} Mannan Ali, Patrick Adie, Christopher H. Marrows, Denis Greig, Bryan J. Hickey and Robert L. Stamps, Nat. Mater. \textbf{6}, 70-75 (2007).

\bibitem{AFM-SG} Eran Maniv, Ryan A. Murphy, Shannon C. Haley, Spencer Doyle, Caolan John, Ariel Maniv, Sanath K. Ramakrishna, Yun-Long Tang, Peter Ercius, Ramamoorthy Ramesh, Arneil P. Reyes, Jeffrey R. Long and James G. Analytis, Nat. Phys.  \textbf{17}, 525-530 (2021). 

\bibitem{Sr2FeCoO6} R. Pradheesh, Harikrishnan S. Nair, V. Sankaranarayanan, K. Sethupathi, Appl. Phys. Lett. \textbf{101}, 142401 (2012).

\bibitem{NiMnIn} B. M. Wang, Y. Liu, P. Ren, B. Xia, K. B. Ruan, J. B. Yi, J. Ding, X. G. Li, and L. Wang, Phys. Rev. Lett. \textbf{106}, 077203 (2011).

\bibitem{Meiklejohn2} W. H. Meiklejohn and C. P. Bean, Phys. Rev. \textbf{105}, 904 (1957).

\bibitem{Review4} R. L. Stamps, J. Phys. D: Appl. Phys. 33, R247 (2000).

\bibitem{FeF2-Fe} J. Nogu\'{e}s, D. Lederman, T. J. Moran, I. K. Schuller, and K. V. Rao, Appl. Phys. Lett. 68, 3186 (1996).

\bibitem{interface_roughness1} A. P. Malozemoff, Phys. Rev. B 35, 3679(R) (1987).

\bibitem{interface_roughness2} T. C. Schulthess and W. H. Butler, Phys. Rev. Lett. 81, 4516 (1998).

\bibitem{Koon} N. C. Koon, Phys. Rev. Lett. 78, 4865 (1997).

\bibitem{minor_loop} A. Harres, M. Mikhov, V. Skumryev, A. M. H. de Andrade, J. E. Schmidt, and J. Geshev, J. Magn. Magn. Mater. 402, 76 (2016).


\bibitem{Hc_TB2} C. Hou, H. Fujiwara, and K. Zhang, Appl. Phys. Lett. 76, 3974–3976 (2000).

\bibitem{Hc_TB1} C. Leighton, J. Nogu\'{e}s, B. J. J\"{o}nsson-{\AA}kerman, and Ivan K. Schuller, Phys. Rev. Lett. 84, 3466 (2000).

\bibitem{positiveEB1} J. Nogu\'{e}s, D. Lederman, T. J. Moran, and Ivan K. Schuller, Phys. Rev. Lett. 76, 4624 (1996).

\bibitem{positiveEB2} S. K. Mishra, F. Radu, H. A. D\"{u}rr, and W. Eberhardt, Phys. Rev. Lett. 102, 177208 (2009).

\bibitem{NiFe-CoO} T. Ambrose and C. L. Chien, Appl. Phys. Lett. 83, 7222-7224 (1998).

\bibitem{Co1-xMgxO} J. Keller, P. Milt\'{e}nyi, B. Beschoten, G. G\"{u}ntherodt, U. Nowak, and K. D. Usadel, Phys. Rev. B 66, 014431 (2002).

\bibitem{Saha} J. Saha and R. H. Victora, Phys. Rev. B 76, 100405(R) (2007).

\bibitem{BiFeO3-Bi2Fe4O9} T. Maity, S. Goswami, D. Bhattacharya, and S. Roy, Phys. Rev. Lett. \textbf{110}, 107201 (2013).

\bibitem{Mn2PtGa} A. K. Nayak, M. Nicklas, S. Chadov, C. Shekhar, Y. Skourski, J. Winterlik, and C. Felser, Phys. Rev. Lett. \textbf{110}, 127204 (2013).

\bibitem{NiMnGa} Z. D. Han, B. Qian, D. H. Wang, P. Zhang, X. F. Jiang, C. L. Zhang, and Y. W. Du, Appl. Phys. Lett. 103, 172403 (2013).

\bibitem{Rev_SG} K. Binder and A. P. Young, Rev. Mod. Phys. 58, 801 (1986).

\bibitem{Murthy} J. Krishna Murthy and A. Venimadhav, Appl. Phys. Lett. \textbf{103}, 25410 (2013).

\bibitem{modelo_PRB} L. T. Coutrim, E. M. Bittar, F. Garcia, and L. Bufai\c{c}al, Phys. Rev. B \textbf{98}, 064426 (2018).

\bibitem{La1.5Ca0.5} L. Bufai\c{c}al, R. Finkler, L. T. Coutrim, P. G. Pagliuso, C. Grossi, F. Stavale, E. Baggio-Saitovitch, and E. M. Bittar, J. Magn. Magn. Mater. \textbf{433}, 271 (2017).

\bibitem{stretched1} R. V. Chamberlin, G. Mozurkewich, and R. Orbach, Phys. Rev. Lett. \textbf{52}, 867 (1984).

\bibitem{stretched2} P. Nordblad, P. Svedlindh, L. Lundgren, and L. Sandlund, Phys. Rev. B \textbf{33}, 645 (1986).

\bibitem{modelo_JMMM} L. Bufai\c{c}al, L. T. Coutrim, E. M. Bittar, F. Garcia, J. Magn. Magn. Mater. 512, 167048 (2020).

\bibitem{Mn3.5Co0.5N} Lei Ding, Lihua Chu, Pascal Manuel, Fabio Orlandi, Meicheng Li, Yanjiao Guo and Zhuohai Liu, Mater. Horiz. 6, 318 (2019).

\bibitem{LaMnO3-PbZr0.8Ti0.2O3} H. J. Mao, C. Song, B. Cui, G. Y. Wang, L. R. Xiao, and F. Pan, J. Appl. Phys. 114, 043904 (2013).

\bibitem{BiFeO3-CoFe2O4} Yue Huang,, Song Li, Zhanqiang Tian, Wenhui Liang, Jing Wang, Xiang Li, Xingwang Cheng, Jun He, Jiping Liu, J. Alloys Compd. 762, 438-443 (2018).

\bibitem{BiFeO3-TbMnO3} Prince K. Gupta, Surajit Ghosh, Shiv Kumar, Arkadeb Pal, Prajyoti Singh, Mohd Alam, Abhishek Singh, Somnath Roy, Rahul Singh, Bheeshma Pratap Singh, N. Naveen Kumar, Eike F. Schwier, Masahiro Sawada, Takeshi Matsumura, Kenya Shimada, Hong-Ji Lin, Yi-Ying Chin, A. K. Ghosh, and Sandip Chatterjee, J. Appl. Phys. 126, 243903 (2019).

\bibitem{MnBi} N. S. Anuraag, S. K. Shaw, Sher Singh Meena, R. K. Singh, N. K. Prasad, J. Magn. Magn. Mater. 557, 169478 (2022).

\bibitem{IrMn-FeCo} A. Migliorini, B. Kuerbanjiang, T. Huminiuc, D. Kepaptsoglou, M. Mu\~{n}oz, J. L. F. Cu\~{n}ado, J. Camarero, C. Aroca, G. Vallejo-Fern\'{a}ndez, V. K. Lazarov and J. L. Prieto, Nature Mater. 17, 28-35 (2018).

\bibitem{Mn3Ga} Linxuan Song, Bei Ding, Hang Li, Senhao Lv, Yuan Yao, Dongliang Zhao, Jun He, Wenhong Wang, J. Magn. Magn. Mater. 536, 168109 (2021).

\bibitem{Mn3-xPtxGa} Ajaya K. Nayak, Michael Nicklas, Stanislav Chadov, Panchanana Khuntia, Chandra Shekhar, Adel Kalache, Michael Baenitz, Yurii Skourski, Veerendra K. Guduru, Alessandro Puri, Uli Zeitler, J. M. D. Coey and Claudia Felser, Nat. Mater. 14, 679-684 (2015).

\bibitem{LaFeO3} Xianke Zhang, Suqiong Xu, Zhiqian Yao, Xiaoqing Liu, Jujun Yuan, Fangguang Kuang, Shuying Kang, Huajun Yu, J. Alloys Compd. 976, 173189 (2024).

\bibitem{JPCM} A. G. Silva, K. L. Salcedo Rodr\'{i}guez, C. P. Contreras Medrano, G. S. G. Louren\c{c}o, M. Boldrin, E Baggio-Saitovitch and L Bufai\c{c}al, J. Phys.: Condens. Matter 33, 065804 (2021).

\bibitem{Mg-ferrite} Himadri Roy Dakua, AIP Adv. 10, 035324 (2020).

\bibitem{Sm1.5Ca0.5CoMnO6} S. K. Giri, R. C. Sahoo, Papri Dasgupta, A. Poddar and T. K. Nath, J. Phys. D: Appl. Phys. 49, 165002 (2016).

\bibitem{NiFe-IrMn} S. K. Mishra, F. Radu, H. A. D\"{u}rr, and W. Eberhardt, Phys. Rev. Lett. 102, 177208 (2009).

\bibitem{Ba3NiIr2O9} Shobha Gondh, Manju Mishra Patidar, Kranti Kumar, M. P. Saravanan, V. Ganesan, and A. K. Pramanik, Phys. Rev. B 104, 014401 (2021).

\bibitem{H_dependence_SG} P. Nordblad, L. Lundgren and L. Sandlund, Europhys. Lett., 3 (2), 235-241 (1987).

\bibitem{LaxSm1-xCrO3} S. Huang, L. R. Shi, Z. M. Tian, H. G. Sun, S. L. Yuan, J. Magn. Magn. Mater. 394, 77-81 (2015).

\bibitem{CoIr} L. T. Coutrim, E. M. Bittar, F. Stavale, F. Garcia, E. Baggio-Saitovitch, M. Abbate, R. J. O. Mossanek, H. P. Martins, D. Tobia, P. G. Pagliuso, and L. Bufai\c{c}al, Phys. Rev. B 93, 174406 (2016).

\bibitem{La1.5Sr0.5Co1-xGaxMnO6} Hongguang Zhang, Wei Chen, Liang Xie, HuiHui Zhao, Qi Li, Curr. Appl. Phys. 35, 58-66 (2022).

\bibitem{Pb6Ni9(TeO6)5} B. Koteswararao, T. Chakrabarty, T. Basu, B. K. Hazra, P. V. Srinivasarao, P. L. Paulose and S. Srinath, Sci. Rep. 7, 8300 (2017).

\bibitem{La1.5Ba0.5} M. Boldrin, L. T. Coutrim, L. Bufai\c{c}al, Braz. J. Phys. 50, 711-715 (2020).

\bibitem{CoIr2} L. T. Coutrim, E. M. Bittar, E. Baggio-Saitovitchb, L. Bufai\c{c}al, J. Magn. Magn. Mater.428, 70-72 (2017).

\bibitem{Ni50Mn38Ga12-xSbx} F. Tian, K. Cao, Y. Zhang, Y. Zeng, R. Zhang, T. Chang, C. Zhou, M. Xu, X. Song and S. Yang, Sci, Rep, 6, 30801 (2016). 

\bibitem{Sr2FeIrO6-La0.67Sr0.33MnO3} K. C. Kharkwal, R. Chaurasia and A. K. Pramanik, J. Phys.: Condens. Matter 31 13LT02 (2019).

\bibitem{SrFeO3-SrCoO3} Tian-Cong Su, Jun Zhang, Wei Zhang, Ying-Ying Wang, Hui-Hui Ji, Xiao-Jiao Wang, Guo-Wei Zhou, Zhi-Yong Quan, Xiao-Hong Xu, Rare Met. 40(7), 1858-1864 (2021).

\bibitem{Bertotti} Giorgio Bertotti, \textit{Hysteresis in Magnetism: For Physicists, Materials Scientists, and Engineers}, Academic Press, San Diego (1998).

\bibitem{SrRuO3} Li Pi, Shixiong Zhang, Shun Tan, and Yuheng Zhang, Appl. Phys. Lett. 88, 102502 (2006).

\bibitem{SrRuO3_comment} Lior Klein, Appl. Phys. Lett. 89, 036101 (2006).

\bibitem{SrRuO3_erratum} Li Pi, Shixiong Zhang, Shun Tan, and Yuheng Zhang, Appl. Phys. Lett. 89, 039902 (2006).

\bibitem{Mydosh} J. A. Mydosh, \textit{Spin Glasses: An Experimental Introduction} (Taylor \& Francis, London, 1993).

\bibitem{CanonicalSG} J. J. Prejean, M. J. Joliclerc, P. Monod, J. Phys. France 41, 427-435 (1980).

\bibitem{La0.7Sr0.3MnO3-NiO} Gyanendra Panchal, R. J. Choudhary, Manish Kumar, D. M. Phase, J. Alloys Compd. 796, 196-202 (2019).

\bibitem{BiFeO3} Sining Dong, Yiping Yao, Ying Hou, Yukuai Liu, Yang Tang and Xiaoguang Li, Nanotechnology 22, 385701 (2011).

\bibitem{SmFeO3} Xiao-xiong Wang, Shang Gao, Xu Yan, Qiang Li, Jun-cheng Zhang, Yun-ze Long, Ke-qing Ruan and Xiao-guang Li, Phys. Chem. Chem. Phys. 20, 3687 (2018).

\bibitem{La1.5Sr0.5Fe0.4Co0.6MnO6} L. Xie, H. G. Zhang, Curr. Appl. Phys. 18, 261-266 (2018).

\bibitem{Co0.8Cu0.2Cr2O4} L. G. Wang, C. M. Zhu, Z. M. Tian, H. Luo, D. L. G. C. Bao, and S. L. Yuan, Appl. Phys. Lett. 107, 152406 (2015).

\bibitem{Mn3.1Sn0.9} Mingyue Zhao, Wei Guo, Xian Wu, Li Ma, Ping Song, Guoke Li, Congmian Zhen, Dewei Zhao and Denglu Hou, Mater. Horiz. 10, 4597-4608 (2023).

\bibitem{Serrate} D. Serrate, J. M. De Teresa, and M. R. Ibarra, J. Phys.: Condens. Matter \textbf{19}, 023201 (2007).

\bibitem{Sami} S. Vasala and M. Karppinen,  Prog. Solid State Chem. \textbf{43}, 1 (2015).

\bibitem{Rev_ASD} Mohd Alam and Sandip Chatterjee, J. Phys.: Condens. Matter 35, 223001 (2023).

\bibitem{Murthy2} J. Krishna Murthy, K. D. Chandrasekhar, H. C. Wu, H. D. Yang, J. Y. Lin and A. Venimadhav, J. Phys.: Condens. Matter 28, 086003 (2016).

\bibitem{La2-xCax} R. C. Sahoo, Sananda Das, Debottam Daw, Ripandeep Singh, A. Das and T. K. Nath, J. Phys.: Condens. Matter 33, 215804 (2021).

\bibitem{La1.5Sr0.5Co1-xFexMnO6} H. G. Zhang, L. Xie, X. C. Liu, M. X. Xiong, L. L. Cao and Y. T. Li, Phys. Chem. Chem. Phys. 19, 25186 (2017).

\bibitem{Er2CoMnO6} A. Banerjee, J. Sannigrahi, S. Giri, and S. Majumdar, Phys. Rev. B 98, 104414 (2018).

\bibitem{Pr2-xSrxCoMnO6} Arkadeb Pal, Prajyoti Singh, V. K. Gangwar, Amish G. Joshi,
P. Khuntia, G. D. Dwivedi, Prince K. Gupta, Mohd Alam, Khyati Anand, K. Sethupathi, Anup K. Ghosh and Sandip Chatterjee, J. Phys.: Condens. Matter 32, 215801 (2020).

\bibitem{Y0.95Eu0.05MnO3} Lixia Xiao, Zhengcai Xia, Zhao Jin, Liran Shi, Yun Ni, Junpei Zhang, Wen Yu, Ceram. Int. 42, 2550-2556 (2016).

\bibitem{PRB2019} L. T. Coutrim, D. Rigitano, C. Macchiutti, T. J. A. Mori, R. Lora-Serrano, E. Granado, E. Sadrollahi, F. J. Litterst, M. B. Fontes, E. Baggio-Saitovitch, E. M. Bittar, and L. Bufai\c{c}al, Phys. Rev. B 100, 054428 (2019).

\bibitem{APL} M. Boldrin, A. G. Silva, L. T. Coutrim, J. R. Jesus, C. Macchiutti, E. M. Bittar, and L. Bufai\c{c}al, Appl. Phys. Lett. 117, 212402 (2020).

\bibitem{meu_La2-xCax} J. R. Jesus, L. Bufai\c{c}al, E. M. Bittar, J. Magn. Magn. Mater. 556, 169402 (2022).

\bibitem{Co-NiMn-filme} A. Akbulut, S. Akbulut, F. Yildiz J. Magn. Magn. Mater. 417, 230-236 (2016).

\bibitem{Calazans} C. Macchiutti, J. R. Jesus, F. B. Carneiro, L. Bufaical, R. A. Klein, Q. Zhang, M. Kirkham, C. M. Brown, R. D. dos Reis, E. M. Bittar, arXiv:2312.09149 (2023).

\bibitem{single_crystal} C. Macchiutti , J. R. Jesus, F. B. Carneiro , L. Bufai\c{c}al, M. Ciomaga Hatnean, G. Balakrishnan, and E. M. Bittar, Phys. Rev. Mater. 5, 094402 (2021).

\bibitem{Ni80Fe20-Mn83Ir17} D. Engel, A. Ehresmann, J. Schmalhorst, M. Sacher, V. H\"{o}ink, G. Reiss, J. Magn. Magn. Mater. 293, 849-853 (2005).

\bibitem{CoO-Ni81Fe19} N. J. G\"{o}kemeijer and C. L. Chien, J. Appl. Phys. 85, 5516 (1999).

\bibitem{tuning_EB} P. Milt\'{e}nyi, M. Gierlings, M. Bamming, U. May, G. G\"{u}ntherodt, J. Nogu\'{e}s, M. Gruyters, C. Leighton, Ivan K. Schuller, Appl. Phys. Lett. 75, 2304-2306 (1999).

\bibitem{Griffiths} Robert B. Griffiths, Phys. Rev. Lett. 23, 17 (1969).

\bibitem{Griffiths2} Wanjun Jiang, XueZhi Zhou, Gwyn Williams, Y. Mukovskii and K. Glazyrin, Phys. Rev. Lett. 99, 177203 (2007).

\end{thebibliography}
\end{document}